\documentclass[prb,showpacs,twocolumn,aps]{revtex4}
  \usepackage[dvipdfm]{graphicx}
\usepackage[dvipdfm]{color}
\usepackage{dcolumn}
\usepackage{amsmath}
\usepackage{amssymb}

\newcommand{\nsno}{$\rm Nd_{1.67}Sr_{0.33}NiO_{4}$}

\newcommand{\lsxno}{$\rm La_{2- \it x}Sr_{\it x}NiO_{4}$}

\newcommand{\lsco}{$\rm     La_{2- \it x}               Sr_{\it x}  Cu                          O_{4}$}
\newcommand{\lbco}{$\rm     La_{2- \it x}  Ba_{\it x}               Cu                          O_{4}$}

\newcommand{\tco}{$T_{\rm CO}$}

\newcommand{\tlt}{$T_{\rm LT}$}
\newcommand{\tcf}{$T_{\rm CF}$}
\newcommand{\ts}{$T_{S}$}

\newcommand{\ets}{$E_{th}^{\rm TS}$}
\newcommand{\etd}{$E_{th}^{\rm D}$}

\begin{document}

\title{Robust charge stripe order under high electric fields in $\bf Nd_{1.67}Sr_{0.33}NiO_4$}

\author{M. H\"ucker$^1$, M. v. Zimmermann$^2$, G. D. Gu$^1$}

\affiliation{$^1$Condensed Matter Physics and Materials Science Department, Brookhaven National Laboratory, Upton, New York 11973, USA}
\affiliation{$^2$Hamburger Synchrotronstrahlungslabor HASYLAB at Deutsches
Elektronen-Synchrotron DESY, 22603 Hamburg, Germany}

\date{\today}

\begin{abstract}
The influence of high electric fields on the charge stripe order in \nsno\ was
studied by means of simultaneous hard x-ray diffraction and electrical
transport experiments. Direct measurements of the charge stripe satellite peaks
in zero and high electric fields provide no evidence for a deformation or a
sliding of the stripe lattice, which contradicts previous indications from
non-linear conductance effects. By using the order parameter of a structural
phase transition for instant sample temperature measurements, non-linear
transport effects can be attributed to resistive heating. Implications for the
pinning of stripes in the nickelates are discussed.
\end{abstract}


\maketitle

%
Electronic ground states with a modulated charge density are a common feature
of correlated electron systems.~\cite{Dagotto05a,Gruener94a} One such class of
materials is represented by the cuprate high temperature superconductors and
isostructural nickelates, such as \lbco\ and \lsxno
.~\cite{Emery93,Tranquada95a} Undoped ($x=0$) these layered
transition-metal-oxides are antiferromagnetic (AF) insulators. When introducing
holes into the $\rm CuO_2$ and $\rm NiO_2$ planes ($x>0$), they become
conducting and the commensurate AF order is
destroyed.~\cite{Kastner98,Tranquada98e} For increasing $x$ the holes
eventually micro phase separate into charge stripes, which form antiphase
boundaries between spin
stripes.~\cite{Tranquada95a,Abbamonte05a,Fujita04a,Kivelson03a} While there is
growing evidence supporting the existence of stripes, their physical properties
and their relevance for the superconductivity in the cuprates are still
unclear. In this context, an important question is how charge stripes respond
to an electric field $E$, whether they start to slide or give rise to non-Ohmic
behavior, and several
experimental~\cite{Yamanouchi99aN,Taguchi00aC,Lavrov03a,Blumberg02aC,
Asamitsu97aM} and theoretical~\cite{Smith98,Hasselmann99a,Bogner01a,Oka03a}
studies have addressed this problem.

A class of correlated electron materials, where non-Ohmic effects are clearly
observed, are one dimensional (1D) charge density wave (CDW) systems, such as
the transition metal chalcogenide $\rm NbSe_3$ and the blue bronze $\rm
K_{0.3}MoO_3$.~\cite{Gruener94a} Here, CDW form as a consequence of Fermi
surface nesting.~\cite{Gruener94a} When the CDW is incommensurate to the
lattice it is usually pinned by impurities or defects such as grain
boundaries.~\cite{Reagor89a,Li99a} A commensurate CDW is expected to couple
stronger to the lattice, although in many cases impurity pinning is still
dominant.~\cite{Dumas83a,Gruener94a} Above an electric threshold field $E_{\rm
th}$, these systems show a strong decrease of the resistivity, associated with
the sliding of a rigid CDW. The threshold $E_{\rm th}$ varies between the
materials and increases with increasing impurity
concentration.~\cite{Reagor89a} Typical values are between 1~mV/cm and a few
V/cm.~\cite{Gruener94a} To gain insight in the sliding process, in several
studies x-rays were used to analyze the CDW satellite reflections as a function
of an applied $E$ field.~\cite{Li99a,Danneau02a,Requardt98a}

In the case of the cuprates and nickelates, significant efforts have been made
to find fingerprints of non-Ohmic transport.~\cite{Yamanouchi99aN,
Taguchi00aC,Lavrov03a} For some systems the results seem to indicate a
dielectric breakdown at much higher thresholds than in classical CDW
materials.~\cite{Yamanouchi99aN,Taguchi00aC} In the nickelates the satellite
reflections associated with the charge stripe order are relatively strong. This
motivated us to directly probe their response to high $E$ fields, and to draw a
comparison to previously reported anomalous transport.~\cite{Yamanouchi99aN}

Here, we present x-ray diffraction and electrical transport measurements on
\nsno , which show that the charge stripes stay locked to the lattice up to $E$
fields on the order of kV/cm. Deviations from Ohm's law have been identified to
result from resistive heating. Crucial to our experiment is the use of an
internal sample-thermometer, which is given by the $T$ dependent super
structure reflection of the low-temperature less-orthorhombic (LTLO)
phase.~\cite{Huecker06a} Results for continuous and pulsed $E$ fields are
qualitatively the same.

\begin{figure}[t]
  \includegraphics [width=0.85\columnwidth,angle=0,clip]{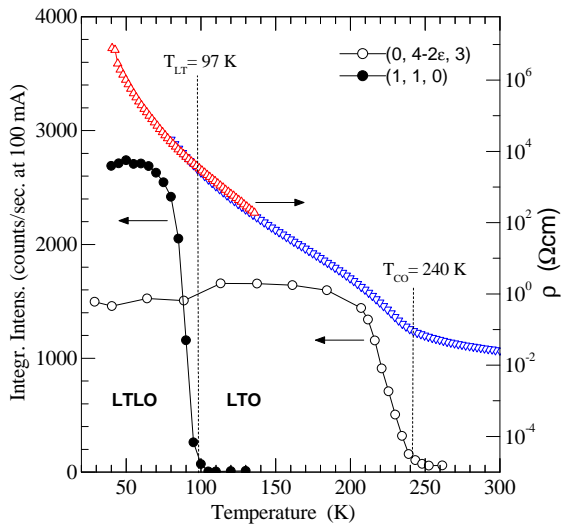}
  \vskip-0.2cm
  \caption{(Color online) $T$ dependence of the in-plane resistivity $\rho$
  as well as the integrated intensity of the charge stripe peak $(0,4-2\epsilon,3)$
  and the super structure peak $(1, 1, 0)$.}\label{fig1}
\end{figure}

Several bar shaped \nsno\ single crystals were prepared with dimensions $0.3
\times 1.2 \times 0.55$~mm$^3$ along the $a$, $b$, and $c$ axis, using the
notation of the low-temperature orthorhombic (LTO) phase.~\cite{Huecker06a} The
in-plane resistivity $\rho$ in Fig.~\ref{fig1} was measured with the four probe
method along the 1.2~mm long direction. Since the crystals are only weakly
twinned, for $\sim$90\% of the sample volume both $a$-axis and charge stripes
are perpendicular to the $E$ field.~\cite{Huecker06a} To measure $\rho$, at low
$T$ (40-135~K) and high $E$ fields a Keithley 2410 source meter was used, and a
Keithley 2000 otherwise. The leads were attached with silver epoxy, carefully
cured to reduce the contact resistance. The sample was mounted on a cold-finger
(CF) using heat conducting epoxy. At low $T$, currents $I \gtrsim 2$~mA
sometimes caused damage to the contacts, so that $I$ was usually limited to
$I_{limit}=1$-1.5~mA. The x-ray studies were performed at beamline BW5 at
HASYLAB. At a photon energy of 100 keV and a beam diameter of $1\times
1$~mm$^2$, the bulk properties of the stripe order and the crystal structure
were studied in transmission geometry.

Figure \ref{fig1} summarizes the structural and transport properties of \nsno\
at zero and very low $E$ fields, respectively. Two structural transitions are
observed. The LTO/LTLO transition at $ T_{\rm LT}\sim$97~K is identified by the
appearance of the $(1,1,0)$ reflection.~\cite{Huecker06a} Apparently, this peak
shows a strong $T$ dependence between 60~K and 100~K, which we have used as an
internal sample thermometer. The charge stripe order below $\sim$240~K is
indicated by satellite reflections with ordering wave vector $q_{CO}=(0,
2\epsilon, 2/3)$, where $\epsilon\simeq x=0.33$, and $1/(2\epsilon)$
corresponds to the in-plane stripe distance.~\cite{Huecker06a}
Figure~\ref{fig1} shows the integrated intensity from $l$ scans through the
charge peak centered at $(0, 4-2\epsilon, 3)$. Below $\sim$180~K the intensity
is constant, except for a step at \tlt , which is also observed for fundamental
reflections, and can be attributed to extinction effects. In $\rho$ the onset
of charge stripe order is marked by a strong increase, whereas the LTO/LTLO
transition leads to no signature (Fig.~\ref{fig1}).

\begin{figure}[t]
  \includegraphics [width=0.7\columnwidth,angle=0,clip]{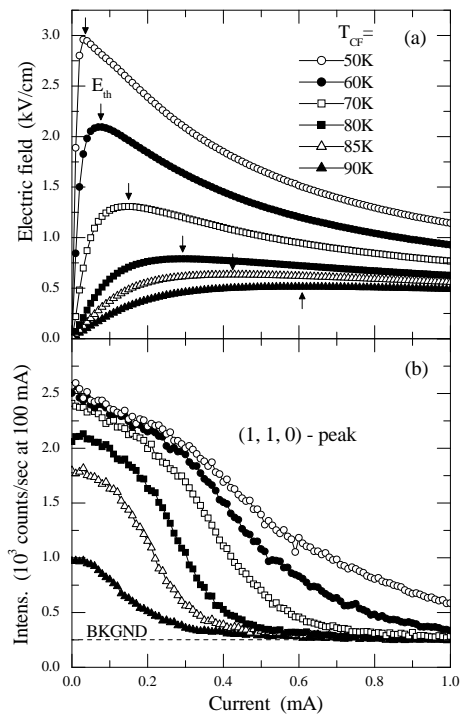}
  \vskip-0.2cm
  \caption{Current dependence of (a) the applied in-plane field $E$
and (b) the $(1, 1, 0)$ peak intensity for different \tcf . The arrows in (a)
mark the threshold field.} \label{fig2}
\end{figure}

Next we turn towards the high $E$ field experiments. In Fig.~\ref{fig2}(a) we
show $I/E$-curves for different cold-finger temperatures \tcf , where $E$ was
measured as a function of a continuous current. The curves are non-linear and
exhibit pronounced maxima. These maxima correspond to the thresholds $E_{\rm
th}$ in an experiment were $I$ is measured as a function of $E$. For $I(E\ll
E_{\rm th})$ the slope $dE/dI$ corresponds to $\rho$ in Fig.~\ref{fig1}.
Simultaneous with the $I$ scans we have measured the $(1,1,0)$ peak intensity
[Fig.~\ref{fig2}(b)]. Apparently, the intensity decreases as $I$ increases,
with the strongest change occurring at lower $I$ for higher $T_{\rm CF}$. Since
the LTO/LTLO transition does not depend on current or voltage, this effect must
have a different reason. To rule out experimental flaws, we have verified that
the (1, 1, 0) peak does not move and that all its intensity is integrated. We
conclude that, with increasing $I$, the sample temperature $T_S$ has increased
significantly above \tcf . In particular, whenever the $(1,1,0)$ intensity
approaches zero, $T_S$ has risen to above $T_{\rm LT}\sim 97$~K. According to
Fig.~\ref{fig1}, an increase of $T_S$ by several 10~K causes $\rho$ to drop by
up to two orders of magnitude, which implies that the non-linear $I/E$-curves
in Fig.~\ref{fig2}(a) are evidence for thermal switching rather than a
dielectric breakdown.

Thermal switching (TS) depends mainly on the thermal coupling between sample
and cold-finger, and the derivative
$d\rho/dT$.~\cite{Kroll74a,Lavrov03a,ZimmermannUnpub} $E_{\rm th}^{\rm TS}$ is
reached when the resistive heating compensates the sample cooling. Once
resistive heating dominates, thermal switching evolves the faster the larger
$-d\rho/dT$. This effect is neither caused by, nor evidence for non-Ohmic
transport. Similar conclusions have been reported for \lsco\ thin films and
$\rm Pr_{0.8}Ca_{0.2}MnO_3$ single crystals.~\cite{Lavrov03a, Mercone05aM}

\begin{figure}[t]
  \includegraphics [width=0.9\columnwidth,angle=0,clip]{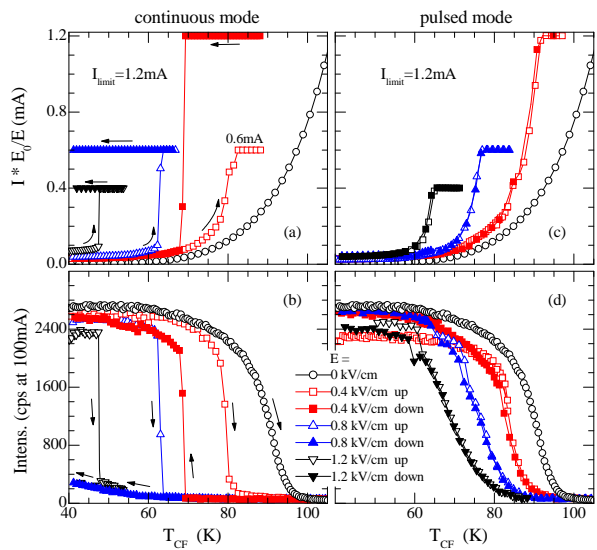}
  \vskip-0.2cm
  \caption{(Color online) In-plane current normalized to $E_0=0.4$~kV/cm,
  and the $(1, 1, 0)$ peak intensity vs \tcf\ for different $E$ fields. (a),(b)
  continuous $E$ field. (c),(d) pulsed
  $E$ field (200~ms). $I_{\rm limit}$ is 1.2~mA (0.6~mA for 0.4~kV up).
  $E$ values are valid for $I<I_{\rm limit}$, only. As soon as $I=I_{\rm limit}$,
  $E$ decreases with increasing \tcf . } \label{fig3}
\end{figure}

To explore if the interfering sample heating can be avoided by means of short
pulses, we have compared $T$ cycles for both pulsed (200~msec) and continuous
$E$ fields. The results in Fig.~\ref{fig3} show the current $I$ and the
$(1,1,0)$ peak intensity as a function of $T_{\rm CF}$ for $E$ fields of
0.4~kV/cm, 0.8~kV/cm, and 1.2~kV/cm. As reference for the case of no sample
heating, a zero field approximation for $I$ was extracted from $\rho(T)$ in
Fig.\ref{fig1}. For continuous $E$ and increasing $T_{\rm CF}$, a sudden
increase of $I$ up to $I_{limit}$ is observed, indicating a drastic decrease of
$\rho$. At the same time the $(1,1,0)$ peak intensity drops to zero
[Figs.~\ref{fig3}(a) and (b)]. This shows that thermal switching has taken
place and that $T_S$ has jumped to a value above $T_{\rm LT}$. Obviously, the
higher $E$ the lower the critical $T_{\rm CF}$ where thermal switching occurs,
which is consistent with the result in Fig.~\ref{fig2}(a) that \ets\ decreases
with increasing \tcf .

Similar results are obtained in the pulsed mode, with 200~msec pulse length and
a 2~sec gap between the pulses [Figs.~\ref{fig3}(c) and (d)]. With increasing
\tcf\ the measured $I$ suddenly starts to increase significantly above the zero
field reference, although not as sharply. Again this effect goes along with a
decrease of the $(1,1,0)$ peak intensity, indicating significant sample
heating. The power dissipated during a 200~msec pulse is of course smaller than
for continuous $E$, which shifts the thermal switching point to somewhat higher
\tcf . Note that photons where counted throughout the entire 200~msec pulse,
whereas $I$ was measured at its end. Therefore, Fig.~\ref{fig3}(d) shows an
average intensity, which is larger than the intensity one would measure during
a very short time at the end of the pulse, if the count rates would allow one
to do so. This explains why the transitions in (1,1,0) are broader than for $I$
and slightly shifted to higher \tcf . When accounting for this effect, the
transitions in Figs.~\ref{fig3}(c) and ~\ref{fig3}(d) become equally sharp and
have the same critical \tcf . For details see
Ref.~\onlinecite{ZimmermannUnpub}.

The continuous mode yields a large \tcf\ hysteresis, which is absent in the
pulsed mode (Fig.~\ref{fig3}). In the pulsed mode no hysteresis is expected,
since the sample cools down to \tcf\ after each pulse; the conditions before a
pulse are the same for the heating and the cooling cycles. In contrast, in the
continuous mode the conditions before each pulse depend on the history. Once
the sample has thermally switched into the low resistive state, \tcf\ must be
decreased significantly before it can return to the high resistive state.
Depending on $E$ and $I_{limit}$, this may not happen even for $T_{\rm
CF}=0$~K.
We conclude that there is no qualitative difference between the pulsed and the
continuous modes. In both cases non-linear conductance effects are caused by
sample heating. In the pulsed mode heating effects are lower, although 200~msec
pulses are still too long to reach significantly higher $E$ fields without
causing a drastic sample heating during the pulse.

\begin{figure}[t]
  \includegraphics [width=0.7\columnwidth,angle=0,clip]{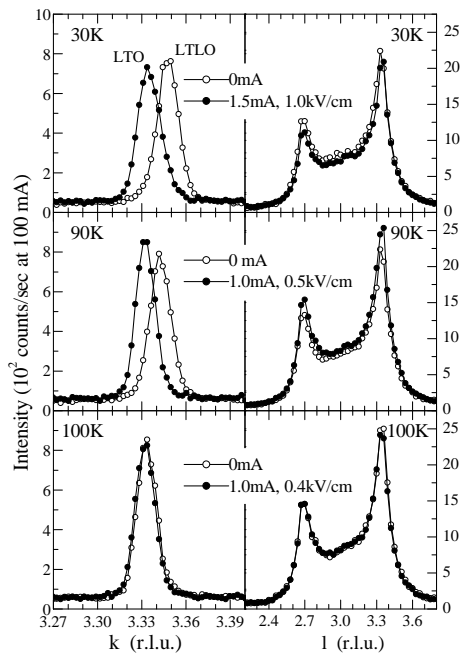}
  \vskip-0.2cm
  \caption{$k$ and $l$ scans through $(0,4-2\epsilon,3)$ charge peak at
  zero and finite continuous current. The peak shift observed in the $k$
  scans is due to the contraction of the $b$ axis in the LTLO phase.} \label{fig4}
\end{figure}

Although our results indicate that in \nsno\ a high $E$ field causes mainly, if
not exclusively, sample heating, it is still possible that it also affects the
stripe order. Therefore, we have studied the shape and intensity of the
$(0,4-2\epsilon,3)$ charge stripe peak for $I=0$~mA and 1~mA (1.5~mA for
$T_{\rm CF}=30$~K), where the finite current is above the threshold for thermal
switching. Figure~\ref{fig4} shows $k$ and $l$ scans for $T_{\rm CF}=30$~K,
90~K and 100~K. The double peak profile of the $l$ scans indicates a
three-layer stripe stacking period.~\cite{Huecker06a}.
At 30~K and 90~K and $I=0$~mA the sample is in the LTLO phase. When turning on
$I$, it transforms into the LTO phase, as is evident from the shift of the peak
in $k$.~\cite{Huecker06a} At $T_{\rm CF}=100$~K the sample is in the LTO phase
for both zero and finite $I$. Obviously, there are no significant differences
in intensity or peak shape between zero and finite $I$ at any \tcf . Both the
in-plane order and the stacking of the stripes are unchanged. Thus, charge
stripes in \nsno\ are robust against $E$ fields on the order of 1~kV/cm. Note
that the slight intensity increase at 90~K and 1~mA is consistent with the
extinction effects at the LTLO/LTO transition (cf. Fig.~\ref{fig1}). At $T_{\rm
CF}=30$~K the slightly smaller intensity at 1.5~mA follows from a peak
broadening in $k$, indicating that $T_S$ is inhomogeneous due to the large
gradient between \ts\ and \tcf .

%
The results in Fig.~\ref{fig4} show that in the studied $T$ region for a
depinning of stripes, if possible at all, higher $E$ fields have to be applied.
This, however, will be difficult because the dissipated energy increases with
$E^2$. As our results show, already for kV/cm fields and mA currents sample
heating is a serious problem (cf. discussion in Ref.~\onlinecite{Lavrov03a}).
For millisecond pulse lengths 10~kV/cm may be reached at low $T$. However, on
this time scale it is difficult to track stripe peak intensities, in addition
to problems emerging from the electrical contacts and thermal stress on the
sample.

It is obvious that the nickelates behave very different from the classical
1D-CDW materials. The stripes emerge from an electronic phase separation,
primarily driven by AF correlations and Coulomb
interactions.~\cite{Tranquada95a} In addition, Hund's-rule exchange causes a
strong localization of the stripes, accompanied by significant local lattice
distortions.~\cite{Anisimov92a,Zaanen94a,Tranquada98a} Hence, each Ni site
constitutes a pinning center. For this type of system it is not surprising that
even for kV/cm electric fields stripes do not slide. Impurities may cause some
disorder, but most likely do not determine the order of magnitude of a
dielectric threshold. Should a dielectric threshold \etd\ exist, it may be
unreachable with conventional $I/E$-experiments. Furthermore, it remains
unclear whether stripes would slide as a whole, break up locally, or
disintegrate as soon as \etd\ is exceeded. The particular scenario depends on
the stiffness of the sliding stripes compared to the pinning forces. Note that
the in-plane correlations of the stripes are stronger than their interplane
correlations. Therefore, we would expect that, with increasing $E$ field, first
the three-layer stacking order degrades, resulting in a change of the double
peak in the $l$ scans. As is evident from Fig.~\ref{fig4}, this is not the case
in our data.

In classical 1D-CDW systems the charge carriers have more itinerant character,
and the electron-lattice coupling is weak. In this case the CDW usually is
pinned by impurities, resulting in a collective mode in the optical
conductivity at energies on the order of $\hbar\omega_0\sim 0.01-0.1$~meV,
which is well below the single particle gap energy.~\cite{Gruener94a} So far,
there is no evidence pointing towards a similar mode in the
nickelates.~\cite{Katsufuji96aN,Homes03aN} Here, the most intriguing feature,
believed to be connected to stripe correlations, is the broad mid-gap infrared
peak at $\sim$0.6~eV.~\cite{Homes03aN} This gap seems to roughly correspond to
the activation energy of $\rho$, indicating that it is most likely associated
with single particle excitations rather than collective
excitations.~\cite{Homes03aN} It also persists to temperatures above \tco
.~\cite{Katsufuji96aN} In a simple model developed for pinned CDW, $ E^D_{\rm
th} \propto \omega_0^2$.~\cite{Gruener94a,Reagor89a} Assuming the case that the
ratio $ E^D_{\rm th} /\omega_0^2$ is the same for 1D-CDW systems and stripes,
this allows one to calculate some rough estimates. On one hand, plugging in
1~kV/cm gives a lower limit for the collective mode of about $\hbar\omega_0
\sim 10$~meV. Although this energy is at least two orders of magnitude larger
than for the mentioned CDW systems, it covers only a small part of the energy
window below the mid-gap peak. On the other hand, it is reasonable to assume
that the energy of any collective mode in the nickelates will not exceed the
mid-gap peak energy. Hence, with $\hbar\omega_0 \sim 0.6$~eV, the upper limit
for \etd\ easily enters the $10^6-10^7$~V/cm range. Unless \etd\ turns out to
be several orders of magnitude smaller, macroscopic non-Ohmic transport in the
stripe ordered nickelates may be out of reach.

In summary, we find that the static charge stripe order in \nsno\ is stable in
electric fields on the order of kV/cm. Neither the in-plane stripe order, nor
the stacking order is affected. Non-linear conductivity effects, previously
associated with a sliding of the stripes, are due to resistive sample heating.
Our results are evidence of the strong charge carrier localization in stripe
ordered nickelates, which distinguishes this class of materials dramatically
from classical CDW systems, where the weakly pinned CDW emerges from Fermi
surface nesting.

We would like to thank J. M. Tranquada, C.~C. Homes and T.~M. Rice for helpful
discussions. The work at Brookhaven was supported by the Office of Science,
U.S. Department of Energy under Contract No.\ DE-AC02-98CH10886.


\begin{thebibliography}{10}

\bibitem{Dagotto05a}
E.\ Dagotto,
\newblock {\em Science} {\bf 309}, 257 (2005).

\bibitem{Gruener94a}
G.~Gr\"uner,
\newblock {\em Density Waves in Solids} volume~89 of {\em Frontiers in
  Physics}.
\newblock Addison-Wesley Publishing Company 1994.

\bibitem{Tranquada95a}
J.~M. Tranquada {\it et al.},
\newblock {\em Nature} {\bf 375}, 561 (1995).

\bibitem{Emery93}
V.J.\ Emery and S.A.\ Kivelson,
\newblock {\em Physica} {\bf {\rm C}~209}, 597 (1993).

\bibitem{Kastner98}
M.~A. Kastner and R.~J. Birgeneau,
\newblock {\em Rev.\ Mod.\ Phys.} {\bf 70}, 897 (1998).

\bibitem{Tranquada98e}
J. M. Tranquada, in
\newblock {\em Neutron Scattering in Layered Copper-Oxide Supercondcutors}, edited by A.~Furrer (Kluwer, Dordrecht, 1998), p. 225.

\bibitem{Abbamonte05a}
P.~Abbamonte {\it et al.},
\newblock {\em Nature Physics} {\bf 1}, 155 (2005).

\bibitem{Fujita04a}
M.~Fujita {\it et al.},
\newblock {\em Phys.\ Rev.} {\bf {\rm B}~70}, 104517 (2004).

\bibitem{Kivelson03a}
S. A.\ Kivelson {\it et al.},
\newblock {\em Rev. Mod. Phys.} {\bf 75}, 1201 (2003).

\bibitem{Yamanouchi99aN}
S.~Yamanouchi {\it et al.},
\newblock {\em Phys.\ Rev.\ Lett.} {\bf 83}, 5555 (1999).

\bibitem{Taguchi00aC}
Y.~Taguchi {\it et al.},
\newblock {\em Phys.\ Rev.} {\bf {\rm B}~62}, 7015 (2000).

\bibitem{Lavrov03a}
A. N.\ Lavrov {\it et al.},
\newblock {\em Phys.\ Rev.} {\bf {\rm B}~68}, 94506 (2003).

\bibitem{Blumberg02aC}
G. Blumberg {\it et al.},
\newblock {\em Science} {\bf 297}, 584 (2002).

\bibitem{Asamitsu97aM}
A. Asamitsu {\it et al.},
\newblock {\em Nature} {\bf 388}, 50 (1997).

\bibitem{Smith98}
C. Morais Smith {\it et al.},
\newblock {\em Phys.\ Rev.} {\bf {\rm B}~58}, 453 (1998).

\bibitem{Hasselmann99a}
N.\ Hasselmann {\it et al.},
\newblock {\em Phys.\ Rev.\ Lett.} {\bf 82}, 2135 (1999).

\bibitem{Bogner01a}
S.\ Bogner and S.\ Scheidl,
\newblock {\em Phys.\ Rev.} {\bf {\rm B}~64}, 54517 (2001).

\bibitem{Oka03a}
T.~Oka {\it et al.},
\newblock {\em Phys.\ Rev.\ Lett.} {\bf 91}, 66406 (2003).

\bibitem{Li99a}
Y.~Li {\it et al.},
\newblock {\em Phys.\ Rev.\ Lett.} {\bf 83}, 3514 (1999).

\bibitem{Reagor89a}
D.~Reagor and G.~Gr\"uner.
\newblock {\em Phys.\ Rev.} {\bf {\rm B}~39}, 7626 (1989).

\bibitem{Dumas83a}
J. Dumas {\it et al.},
\newblock {\em Phys.\ Rev.\ Lett.} {\bf 50}, 757 (1983).

\bibitem{Danneau02a}
R.~Danneau {\it et al.},
\newblock {\em Phys.\ Rev.\ Lett.} {\bf 89}, 106404 (2002).

\bibitem{Requardt98a}
H.~Requardt {\it et al.},
\newblock {\em Phys.\ Rev.\ Lett.} {\bf 80}, 5631 (1998).

\bibitem{Huecker06a}
M.\ H\"ucker {\it et al.},
\newblock {\em Phys.\ Rev.} {\bf {\rm B}~74}, 85112 (2006).

\bibitem{Kroll74a}
D.M. Kroll,
\newblock {\em Phys.\ Rev.} {\bf {\rm B}~9}, 1669 (1974).

\bibitem{ZimmermannUnpub}
M.~v.\ Zimmermann {\it et al.}, (unbuplished).

\bibitem{Mercone05aM}
S.~Mercone {\it et al.},
\newblock {\em J. Appl. Phys.} {\bf 98}, 23911 (2005).

\bibitem{Anisimov92a}
V.~I.\ Anisimov {\it et al.},
\newblock {\em Phys.\ Rev.\ Lett.} {\bf 68}, 345 (1992).

\bibitem{Zaanen94a}
J.\ Zaanen and P.~B.\ Littlewood,
\newblock {\em Phys.\ Rev.} {\bf {\rm B}~50}, 7222 (1994).

\bibitem{Tranquada98a}
J.~M.\ Tranquada,
\newblock {\em J.\ Phys.\ Chem.\ Solids} {\bf 59}, 2150 (1998).


\bibitem{Katsufuji96aN}
T.~Katsufuji {\it et al.},
\newblock {\em Phys.\ Rev.} {\bf {\rm B}~54}, 14230 (1996).

\bibitem{Homes03aN}
C.C. Homes {\it et al.},
\newblock {\em Phys.\ Rev.} {\bf {\rm B}~67}, 184516 (2003).

\end{thebibliography}


\end{document}